\documentclass[aps,prl,twocolumn,groupedaddress]{revtex4}  

\usepackage{graphicx}
\usepackage{dcolumn}
\usepackage{bm}
\pdfoutput=1

\begin{document}


\title{Photoluminescence from In$_{0.5}$Ga$_{0.5}$As/GaP quantum dots coupled to photonic crystal cavities}

\author{Kelley Rivoire$^{1}*$, Sonia Buckley$^{1}$, Yuncheng Song$^{2}$, Minjoo Larry Lee$^{2}$, Jelena Vu\v{c}kovi\'{c}$^{1}$}

\email{krivoire@stanford.edu} 

\address{$^1$E. L. Ginzton Laboratory, Stanford University, Stanford, CA 94305-4085\\ $^2$Department of Electrical Engineering, Yale University, New Haven, CT, 06511}

\begin{abstract}
We demonstrate room temperature visible wavelength photoluminescence from In$_{0.5}$Ga$_{0.5}$As quantum dots embedded in a GaP membrane. Time-resolved above band photoluminescence measurements of quantum dot emission show a biexpontential decay with lifetimes of $\approx$200 ps. We fabricate photonic crystal cavities which provide enhanced outcoupling of quantum dot emission, allowing the observation of narrow lines indicative of single quantum dot emission. This materials system is compatible with monolithic integration on Si, and is promising for high efficiency detection of single quantum dot emission as well as optoelectronic devices emitting at visible wavelengths.
\end{abstract}

\maketitle

Semiconductor quantum dot (QD) emitters grown in gallium phosphide are important for both classical optoelectronic and quantum applications. The close match between the lattice constants of GaP and Si (0.37\% at 300K \cite{Grassman}) is promising for monolithic integration with silicon\cite{Grassman,lin,talebi}, and the large electronic band gap of GaP allows light emission at visible wavelengths. Single quantum dots (QDs) at visible wavelengths are beneficial for quantum applications since Si avalanche photodiodes (APDs) have maximum quantum efficiency in the red part of spectrum; additionally, emission in this part of the spectrum can be frequency downconverted to telecommunications wavelengths using readily available lasers\cite{curtz, zaske}.

Quantum dots emitting in the red have been extensively studied over the past decade in materials systems including InP/InGaP\cite{ugur_apl,luxmoore, richter, reischle}, InP/GaP\cite{hatami_prb, hatami_apl_2001}, InP/AlGaInP\cite{schulz, schulz_algainp}, GaInP/GaP\cite{gerhard_2009}, InAs/GaP\cite{leon}, and AlGaInP/GaP\cite{gerhard}. Of these systems, clear single quantum dots with narrow emission lines exhibiting antibunching have been observed only in the InP/InGaP and InP/AlGaInP systems. GaP-based materials, by contrast, allow either monolithic integration with Si or growth on a non-absorbing GaP substrate (due to the large indirect electronic band gap); additionally, the stronger second order optical nonlinearity of GaP compared to InGaP is preferable for on-chip frequency downconversion to telecom wavelengths. Recently\cite{song}, low temperature emission (80K) was measured from In$_{0.5}$Ga$_{0.5}$As self-assembled QDs in GaP emitting in the red part of the spectrum. This system provides large wavelength tunability as the In fraction can be varied from 0.07-0.50 without introducing dislocations; additionally it should provide deeper confinement for carriers than InP/GaP. Subsequently, room temperature emission was measured from In$_{0.3}$Ga$_{0.7}$As/GaP QDs\cite{tranh}; measured temperature dependence of emission and supporting tight binding calculations indicated good confinement of carriers and type-I emission. Here, we further characterize this materials system, integrate it with photonic nanostructures that enhance the emission of the QDs, and observe evidence indicative of emission from individual QDs.

\begin{figure}[h]
\includegraphics[width=8.5cm]{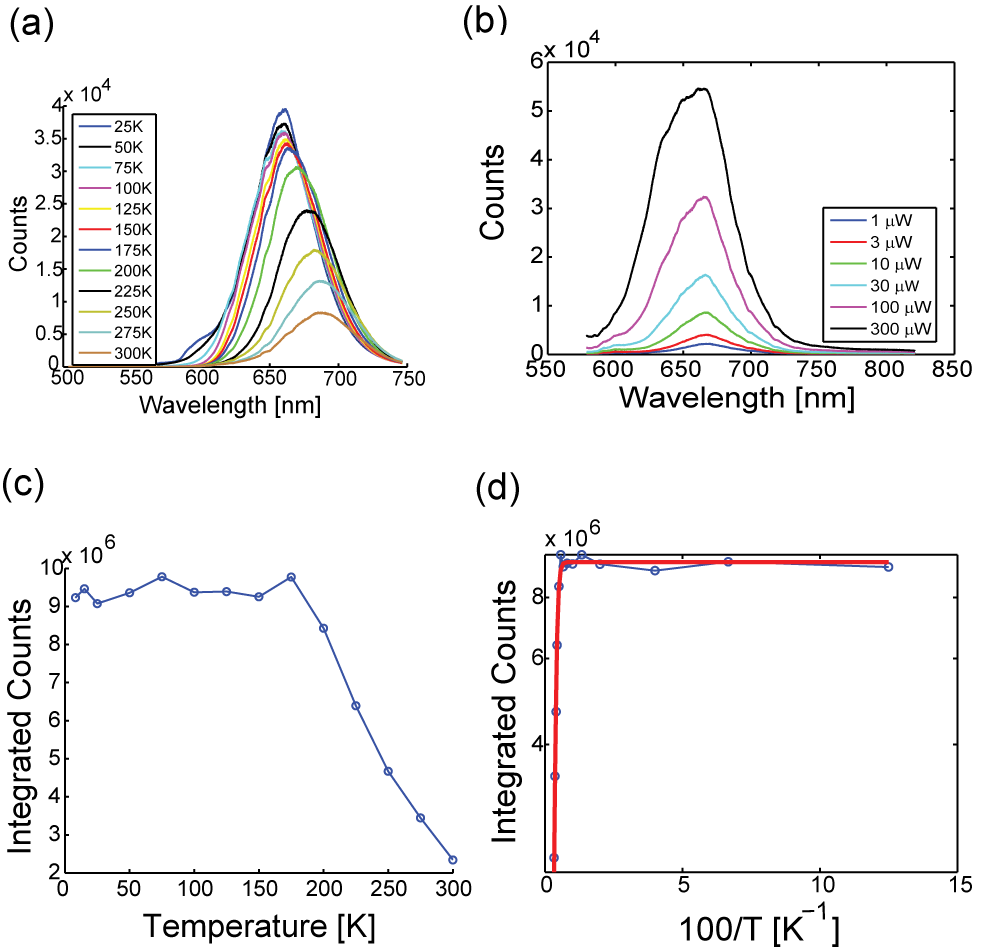}
\caption{\label{fig:figure1}
    (a) QD PL from an unprocessed region of sample as temperature is varied between 25K and 300K. The pump is a 405 nm CW laser diode at 700 $\mu$W. (b) Photoluminescence as a function of pump power at 10K showing broadening at higher energies with increasing power, indicative of the presence of excited states. (c) Integrated counts of low energy half of PL spectrum (to minimize contribution of excited states, which varies with temperature). Intensity is decreased by a factor of 4 at 300K. (d) Semilog plot of integrated counts of low energy half of PL spectrum versus inverse temperature. Fit to exponential (red line) gives activation energy $E_a$=161 meV.}
\end{figure}

The QDs are grown by solid source molecular beam epitaxy in the center of a 200 nm thick GaP membrane grown on top of a 500 nm layer of Al$_{0.8}$Ga$_{0.2}$P on a (001) GaP substrate. Fig. 1a shows the measured QD photoluminescence (PL) as a function of temperature from 25K to 300K in a continuous flow helium cryostat using 700 $\mu$W excitation power from a 405 nm continuous wave (CW) diode laser. (The power level was chosen to maintain a constant integration time on the spectrometer CCD for all temperatures.) The center wavelength of the QD emission redshifts by 30 nm from 25K to 300K. The large full width half maximum of the emission is expected to result from inhomogeneous broadening due to variation in the physical size of the QDs\cite{song}. Fig. 1b shows that for higher pump powers, the QD spectrum broadens on the high energy side and the integrated intensity is nonlinear as a function of power, indicating the presence of excited states. To characterize the confinement of carriers in the QDs, we study the intensity of QD emission as a function of temperature. Fig. 1c shows the emission intensity integrated across the low energy half of QD PL spectrum (to minimize the contribution of excited states) from Fig. 1a. The emission intensity decreases by a factor of 4 from cryogenic temperatures to room temperature. Fig. 1d shows a fit of this integrated intensity to an Arrenhius model (assuming a temperature-independent radiative lifetime) with a single activation energy with form:
\begin{equation}
\frac{I(T)}{I_0} = \frac{1}{1+C\times\exp\frac{-E_a}{kT}}
\end{equation}
where $I(T)$ is the temperature-dependent intensity, $I_0$ is the intensity at 0K, $C$ is a constant, $k$ is Boltzmann's constant, and $E_a$ is the activation energy indicating carrier confinement. We measure $E_a$=161 meV; this is slightly larger than the 134 meV measured by Tranh et al\cite{tranh}, most likely due to the larger indium content in our samples, which results in deeper confinement.

\begin{figure}[h]
\includegraphics[width=8.5cm]{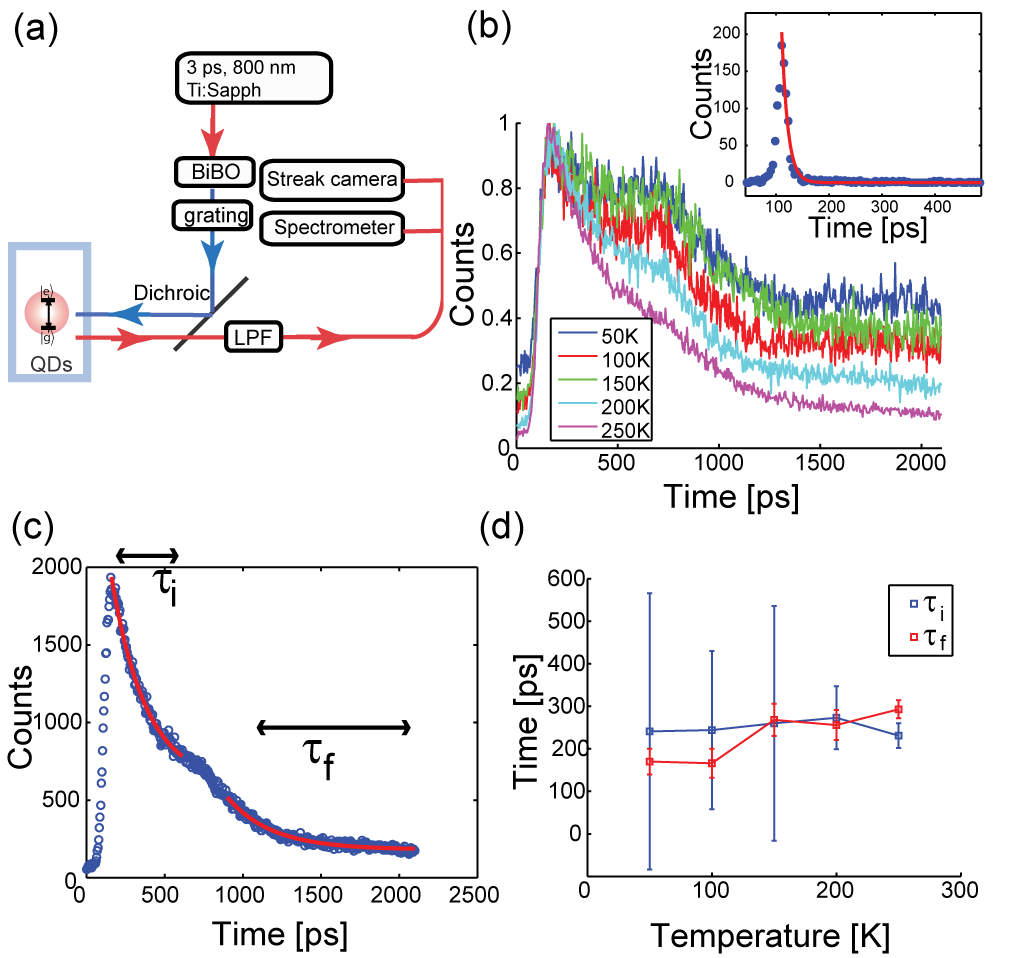}
\caption{\label{fig:figure1}
    (a) Experimental setup for time-resolved measurements. 3 ps pulses at 800 nm from Ti:Sapphire laser are frequency doubled in a BiBO crystal. A grating is used to filter the 400 nm light, which is passed through a dichroic mirror onto the sample. The photoluminescence emitted by the quantum dots is transmitted through the dichroic, passed through a long-pass filter (LPF) to remove any residual pump, and sent to a spectrometer or streak camera (for time-resolved measurements). (b) Time-resolved streak camera measurements showing lifetime integrated across low energy half of PL spectrum for different temperatures. Inset: excitation pulse at 400 nm. Red line indicates fit with decay time 12 ps, limited by instrument resolution. (c) Exponential fits of time-resolved data at 250K, showing initial and final decay times $\tau_i$ and $\tau_f$. (d) Measured lifetimes as a function of sample temperature. }
\end{figure}

We investigate the dynamics of the ensemble QD emission by studying the time-resolved photoluminescence on a streak camera when the quantum dots are excited at 400 nm by a frequency doubled Ti:Sapphire laser with a repetition rate of 80 MHz. The experimental setup is shown in Fig. 2a; the instrument response to the pump (12 ps) is shown in Fig. 2b inset. The time-resolved emission of the low energy half of the QD spectrum (Fig. 2b) for all temperatures shows a biexponential decay with a short component ($\tau_i\approx$ 250 ps, averaged across all temperatures) followed by a decay with similar time constant ($\tau_f\approx$ 230 ps) after a delay of $\approx$500 ps. The delay is most likely indicative of phonon-assisted
transfer of carriers from the indirect GaP matrix\cite{hatami_prb}. Fig. 2c shows the time-resolved PL measured at 250K, indicating regions used for fitting initial and final time constants. The extracted time constants for each temperature  are shown in Fig. 2d; error bars indicate error from fit. The short lifetime is consistent with a Type I system; the minimal temperature dependence of decay rates indicates the absence of temperature-dependent non-radiative processes.

\begin{figure}[h]
\includegraphics[width=8.5cm]{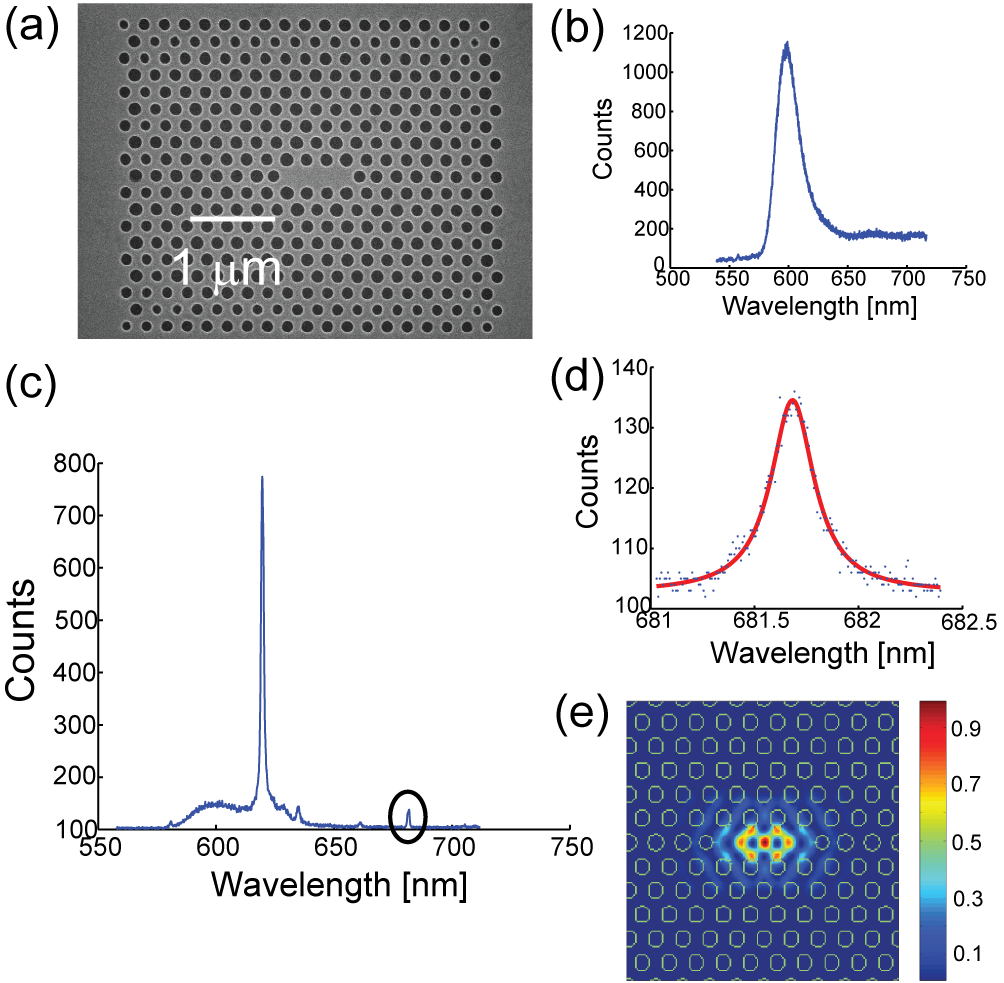}
\caption{\label{fig:figure2}
    (a) SEM image of photonic crystal nanocavity. (b) PL measured at 12K with 405 nm CW pump from unprocessed region of thinner sample used for photonic crystal measurements. QD wavelength is slightly blueshifted from Fig. 1. (c) PL measurement indicating emission of quantum dots coupled into cavity modes. Fundamental cavity mode is indicated by black circle. (d) Lorenztian fit of fundamental cavity mode at 681.7 nm with Q=2800. (e) FDTD-simulated electric field intensity for fundamental cavity mode.}
\end{figure}

To fabricate photonic crystal cavities, we used a different sample with a thinner 93 nm-thick GaP membrane for ease in fabrication. The photonic crystals were fabricated by top-down fabrication, including e-beam lithography, dry etching, and wet etching to remove the sacrificial AlGaP layer\cite{rivoire_gap}. A scanning electron microscope (SEM) image of a fabricated sample is shown in Fig. 3a. Due to a difference in MBE growth parameters, the emission wavelength of the thinner sample was slightly blueshifted, as shown in Fig. 3b, and the QD density was lower. Fig. 3c shows photoluminescence, measured at 12K, from the quantum dots coupled into the linear three-hole defect photonic crystal cavity\cite{noda}. The fundamental mode of the cavity with highest quality factor (black circle) overlaps with the tail of the QD emission; the brighter higher order cavity modes\cite{chalcraft} are more closely matched to the QD emission spectrum. Fig. 3d shows the fundamental cavity mode resonance at 681.7 nm; a Lorentzian fit indicates a quality factor of 2800; Fig. 3e shows a finite difference time domain (FDTD) simulation of the spatial distribution of the electric field intensity in the center of the membrane for this cavity mode.

\begin{figure}[h]
\includegraphics[width=8.5cm]{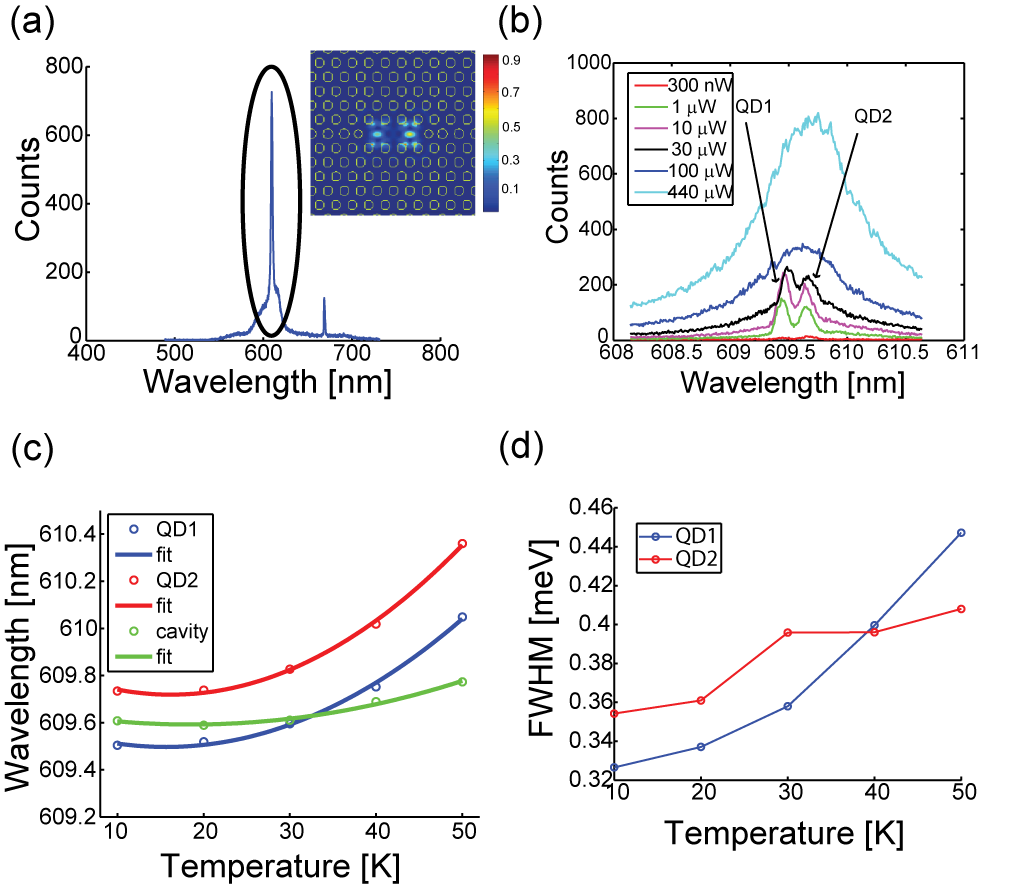}
\caption{\label{fig:figure3}
   (a) PL from a different photonic crystal cavity structure with higher order cavity mode aligned to wavelength of strongest QD PL measured with 405 nm CW pump. Black circle indicates mode of interest. Inset: FDTD-simulated electric field intenstiy for higher order cavity mode indicated by circle in (a). (b) High resolution spectra of QD PL from cavity in (a). Measurements are taken at 10K. The two single lines at low power are indicative of single quantum dots. (c)
  Change in wavelength of quantum dots (measured at 500 nW) and cavity (measured at 100 $\mu$W) as a function of temperature. Solid lines indicate quadratic fits. (d) Change in QD linewidth as a function of temperature.}
\end{figure}

By measuring the cavity-enhanced QD emission, we are able to observe indications of single quantum dot emission lines at temperatures below 60K. Fig. 4a shows the low temperature PL spectrum with a CW pump from another photonic crystal cavity on the chip. The black circle indicates the higher order mode of interest which we use to enhance the quantum dot emission outcoupling. The inset shows the spatial profile of electric field intensity for the mode of interest, calculated by the FDTD method. Fig. 4b shows a high resolution spectrum measured at 10K of the QD and cavity mode at ~610 nm (black circle in Fig. 4a). At low power, two narrow lines appear to saturate (as would be expected for single QDs) as power is increased above about 30 $\mu W$. At higher powers, we recover the Lorentzian lineshape of the photonic crystal cavity mode, as the intensity in the cavity mode continues to grow while intensity from the individual QD lines has saturated.
Fig. 4c shows the wavelength shift of quantum dots and cavity as a function of temperature, indicating a quadratic redshift in dot emission as temperature is increased, as expected due to the approximately quadratic shift in material band gaps in this temperature range\cite{varshni}. Quantum dot lines are measured at 500 nW (far below QD saturation), while the cavity is measured at 100 $\mu$W (above QD saturation). The quantum dot emission wavelength changes at a different rate than the cavity emission, confirming that the narrow lines are not associated with a cavity mode. Fig. 4d shows the change in full width half max (FWHM) of the observed spectral lines at 1 $\mu$W power as a function of temperature measured. The narrow QD-like lines show an increase in linewidth as the temperature is raised, as expected for single quantum dots, while the cavity linewidth remains roughly unchanged in the same temperature range. Further confirmation of single QD behavior could be obtained from photon statistics measurements. We did not obtain sufficient signal-to-noise from the cavity to perform such measurements in this case. An improvement in the signal-to-noise, for example by improving the cavity quality factor, would also allow an investigation of the time-resolved dynamics of a single QD coupled to the cavity, where Purcell enhancement is expected\cite{dirk_prl} in this regime\cite{dirk_prl}.

In conclusion, we measure the temperature-dependent photoluminescence from In$_{0.5}$Ga$_{0.5}$As quantum dots embedded in a GaP membrane, indicating good carrier confinement with only a four-fold decrease in emitted intensity from cryogenic to room temperature. We study the temperature-dependent time-resolved photoluminescence, which shows a biexponential decay with time constants of $\approx$ 200 ps. We also observe enhanced emission into the modes of a photonic crystal cavity and narrow lines consistent with single quantum dot emission. The materials system is compatible with monolithic integration on Si and is also promising for quantum applications. The quantum dot wavelength is matched to the high efficiency region of silicon APDs and could be downconverted to telecommunication wavelengths through integration with photonic nanostructures\cite{shg_rivoire_opex, rivoire_wg_shg, rivoire_sfg, mccutcheon}.

This work was supported by the National Science Foundation, a National Science Graduate Fellowship, and Stanford Graduate Fellowships. MLL acknowledges the DARPA Young Faculty Award program (Grant No. N66001-11-1-4148).

\end{document}